\documentclass[conference]{IEEEtran}
\IEEEoverridecommandlockouts
\usepackage{cite}
\usepackage[size=smaller]{caption}
\usepackage{amsmath,amssymb,amsfonts, bbm}
\usepackage{subcaption}
\usepackage{stfloats} 
\usepackage{booktabs}
\usepackage{graphicx}
\usepackage{textcomp}
\usepackage{xcolor}
\usepackage{array}

\usepackage[ruled,vlined]{algorithm2e}
\usepackage{multicol}

\def\BibTeX{{\rm B\kern-.05em{\sc i\kern-.025em b}\kern-.08em
    T\kern-.1667em\lower.7ex\hbox{E}\kern-.125emX}}
\begin{document}

\title{
Input-Specific and Universal Adversarial Attack Generation for Spiking Neural Networks in the Spiking Domain\thanks{This work was supported by the French National Research Agency (ANR) through the European CHIST-ERA program under the project TruBrain (Grant N$^{\mbox{\scriptsize o}}$ ANR-23-CHR4-0004-01) and by the European Network of Excellence dAIEDGE (Grant  N$^{\mbox{\scriptsize o}}$ 101120726).}
}

\author{
    \IEEEauthorblockN{
        Spyridon Raptis and Haralampos-G. Stratigopoulos}
    \IEEEauthorblockA{\small{
        Sorbonne Université, CNRS, LIP6, Paris, France}}}

\maketitle

\begin{abstract}
As Spiking Neural Networks (SNNs) gain traction across various applications, understanding their security vulnerabilities becomes increasingly important. In this work, we focus on the adversarial attacks, which is perhaps the most concerning threat. An adversarial attack aims at finding a subtle input perturbation to fool the network's decision-making. We propose two novel adversarial attack algorithms for SNNs: an input-specific attack that crafts adversarial samples from specific dataset inputs and a universal attack that generates a reusable patch capable of inducing misclassification across most inputs, thus offering practical feasibility for real-time deployment. The algorithms are gradient-based operating in the spiking domain proving to be effective across different evaluation metrics, such as adversarial accuracy, stealthiness, and generation time. Experimental results on two widely used neuromorphic vision datasets, NMNIST and IBM DVS Gesture, show that our proposed attacks surpass in all metrics all existing state-of-the-art methods. Additionally, we present the first demonstration of adversarial attack generation in the sound domain using the SHD dataset.
\end{abstract}


\section{Introduction} \label{sec:introduction}

Neuromorphic computing has emerged as a revolutionary paradigm, inspired by the structure and functionality of the human brain, to address the growing demand for low-power and low-latency inference in Artificial Intelligence (AI). Neuromorphic computing is based on Spiking Neural Networks (SNNs), which represent a biologically plausible model of neural computation, leveraging discrete spikes to process information. SNNs have shown significant promise in various applications, such as vision, speech processing, and robotics \cite{RoJaPa19, SKPMDK22}.

With the growing reliance on AI systems, unique security risks and challenges emerge, underscoring the need to understand the vulnerabilities and threat landscape in neuromorphic computing as the field continues to evolve rapidly. Perhaps the most preoccupying threat is the adversarial attack \cite{SZSBEG14}, which aims at exploiting subtle perturbations in the input to manipulate the network’s behavior, leading to incorrect predictions. While adversarial attacks have been studied extensively for Artificial Neural Networks (ANNs) \cite{CRPI24}, transferring these attacks to SNNs presents unique challenges due to their discrete and temporal characteristics. 

Adversarial attacks on SNNs can be classified in various ways. They presume a real-valued input source, i.e., image frames, which is then encoded into spike trains \cite{comprehensiveAnalysysAdversarial,IsSpikingSecure,sharmin2020inherent, SecuringDeepSNNs}, or they operate directly on spiking datasets, for example extracted by a Dynamic Vision Sensor (DVS) \cite{BaSiRa18,marchisio2021dvsattacksadversarialattacksdynamic,spikeFool,AdversarialG2S}. Although spike encoding bridges the gap between conventional data and SNNs, SNNs are inherently designed to work with spiking input datasets because such datasets align naturally with their event-driven, time-based computational paradigm. This is where SNNs showcase their advantages, as opposed to converting static frames to spiking data, which results in energy cost, loss of temporal richness, and reduced biological fidelity. Therefore, techniques designed for spiking input sources \cite{BaSiRa18,marchisio2021dvsattacksadversarialattacksdynamic,spikeFool,AdversarialG2S} are necessary to accommodate the most efficient and natural application of SNNs.

Another categorization of existing techniques is trial-and-error \cite{IsSpikingSecure,BaSiRa18,marchisio2021dvsattacksadversarialattacksdynamic} versus gradient-based \cite{sharmin2020inherent,SecuringDeepSNNs,marchisio2021dvsattacksadversarialattacksdynamic,spikeFool,AdversarialG2S}. Trial-and-error refers to an iterative process where the input sample is repeatedly modified until the desired adversarial behavior is achieved. This approach is not effective because it is difficult to find the optimal perturbation in the limited search time that the attacker has.

Gradient-based approaches rely on the model’s internal gradients to quickly determine the optimal perturbation. When applied to SNNs \cite{marchisio2021dvsattacksadversarialattacksdynamic,spikeFool,AdversarialG2S}, each iteration of the algorithm is composed of three stages: (i) forward pass to obtain the model prediction and compute the loss; (ii) backward pass to compute the gradient of the loss with respect to the input of the first layer; and (iii) update of the spiking input perturbation. The backward pass is performed in the spiking domain using surrogate gradients to address the non-differentiability of the spiking activation function \cite{WDLZS18,shor18}. The proposed techniques mainly differ in step (iii), in particular how the continuous gradients at the input of the first layer, which are typically with respect to the membrane potential of the neurons, are converted to perturbations in the spiking domain, with the challenge being to maintain the spatiotemporal gradient information during the conversion. 

A third categorization is input-specific versus universal adversarial attacks. Input-specific attacks generate perturbations that are specifically crafted for each individual input. In contrast, universal attacks generate a single perturbation that can fool the model across many inputs. Input-specific attacks are less plausible as it is assumed that the attacker can capture the running input, generate the adversarial version offline, and modify the input, all in real-time without introducing any delay. Universal attacks are more practical and efficient in real-world scenarios. The vast majority of works propose input-specific attacks \cite{comprehensiveAnalysysAdversarial,IsSpikingSecure,sharmin2020inherent, SecuringDeepSNNs,BaSiRa18,marchisio2021dvsattacksadversarialattacksdynamic, spikeFool,AdversarialG2S} with only one work addressing the universal attack problem \cite{spikeFool}. 

Metrics used to evaluate the efficiency of adversarial attacks include: (i) adversarial accuracy or attack success rate (ASR) defined as the percentage of adversarial examples that successfully cause the model to produce an incorrect answer; (ii) perturbation size defined as the percentage of input spikes across the complete input duration that are flipped; and (iii) average elapsed time to generate the adversarial example.

In this work we make the following contributions:

\begin{enumerate}
    \item We propose novel gradient-based input-specific and universal adversarial attacks that operate exclusively in the spiking domain, effectively preserving the spatiotemporal information within the gradients. 
    \item Our input-specific attack outperforms all prior state-of-the-art gradient-based attacks in the spike domain in all metrics \cite{marchisio2021dvsattacksadversarialattacksdynamic,spikeFool,AdversarialG2S}.
    \item Our universal attack outperforms the only existing approach \cite{spikeFool} as it is more successful across the entire dataset and is inherently more stealthy.
    \item While previous works focus on vision datasets \cite{marchisio2021dvsattacksadversarialattacksdynamic,spikeFool,AdversarialG2S}, such as the NMIST \cite{OJCT15} and IBM DVS Gesture \cite{ATBM17}, we demonstrate for the first time adversarial attacks on the Spiking Heidelberg Digits (SHD) sound dataset \cite{heidelberg}.
\end{enumerate}

The rest of the article is structured as follows. In Section \ref{sec:state_of_the_art}, we review the state-of-the-art on adversarial attacks on SNNs following the aforementioned categorization. In Section \ref{sec:snn_input_nature}, we discuss the generation of spiking datasets. In Section \ref{sec:algorithms}, we describe the proposed adversarial attacks generation. In Section \ref{sec:exp_setup}, we describe the experimental setup and, in Section \ref{sec:results}, we present the results. Section \ref{sec:conclusion} concludes the paper by pointing to future work ideas.

\section{State-of-the-art adversarial attacks on SNNs}\label{sec:state_of_the_art}

\subsection{Input-specific adversarial attacks}

\subsubsection{Real-valued input}

In \cite{comprehensiveAnalysysAdversarial}, an ANN model with the same topology as the SNN is randomly initialized and its weights are overwritten with the weights of the SNN. Next, the real-valued input is converted into the spiking domain using Poisson rate encoding and converted back to a real-valued rate input by averaging the spikes for each pixel over the sample duration. The classical Fast Gradient Sign Method (FGSM) \cite{FGSM2015} is used to generate an adversarial input for the ANN starting from the rate input. The ANN adversarial input is then converted to the adversarial SNN input using the same Possion rate coding. 

In \cite{IsSpikingSecure}, the algorithm follows a trial-and-error approach applying a perturbation in a subset of pixels in the middle of the image that is imperceptible to the human eye so as to maximize the difference between the target class probability and the maximum class probability considering all other classes. 

In \cite{sharmin2020inherent}, the classical adversarial generation algorithms FGSM \cite{FGSM2015} and Projected Gradient Descent (PGD) \cite{PGD2019} proposed for ANNs are adapted in the spiking domain. A similar approach is used in \cite{SecuringDeepSNNs} using PGD, but the focus was to study how the spiking structural parameters, such as neuron threshold and inference window, can affect the adversarial robustness.

\subsubsection{Spiking input}

In \cite{BaSiRa18}, an adversarial input is generated by adding or removing spikes in a trial-and-error fashion, under the constraint that the number of perturbed spikes stays confined within a fraction $\epsilon$ of the number of input spikes.

In \cite{marchisio2021dvsattacksadversarialattacksdynamic}, both trial-and-error and gradient-based attacks are proposed for vision datasets generated by a DVS. Trial-and-error attacks perturb spikes in the perimeter of the frame (\textit{frame} attack), in the corners (\textit{corner} attack) or in two pixels at a time (\textit{dash} attack) for the whole duration of the sample. The gradient-based attack, called \textit{sparse} attack, works by updating the input perturbation based on the gradient of the loss with respect to the input.

In \cite{spikeFool}, \textit{SpikeFool} is proposed, which is an adapted version of the \textit{SparseFool} attack \cite{MM-DF19} in the spiking domain. \textit{SparseFool} finds a small perturbation on the input in the direction orthogonal to the decision boundary. 

In \cite{AdversarialG2S}, a gradient-based attack is proposed specific to SNNs that leverages the spatiotemporal backpropagation SNN training \cite{WDLZS18}.
In the backward pass, the real-valued gradients with respect to the membrane potential of the input neurons is generated. Then, a gradient-to-spike (G2S) converter using a series of mathematical operations is proposed to convert the first-layer input gradient map to input spiking train updates in each iteration. For some inputs, the gradient vanishing problem was encountered, i.e., the resultant gradient map had all zeros. For such inputs, to circumvent this issue, the restricted spike flipper (RSF) converter is proposed that re-initializes the gradient map and is combined with increasing the firing rate of neurons in the penultimate layer during the attack.

\subsection{Universal adversarial attacks}

In \cite{spikeFool}, the universal patch is restricted in the area of the input that is key for correct prediction. For example, in the case of a vision input, it is restricted in the area where the action is performed. The proposed algorithm is adapted from \cite{BMRAG18}. The patch is optimized iteratively on the inputs using the PGD gradient-based method to maximize the misclassification rate across inputs.
\section{Spiking datasets}\label{sec:snn_input_nature}

Spiking inputs are binary signals that represent discrete events over time, mimicking the way biological neurons communicate with one another.

\begin{figure}[t]
\centering
\includegraphics[width=0.8\linewidth]{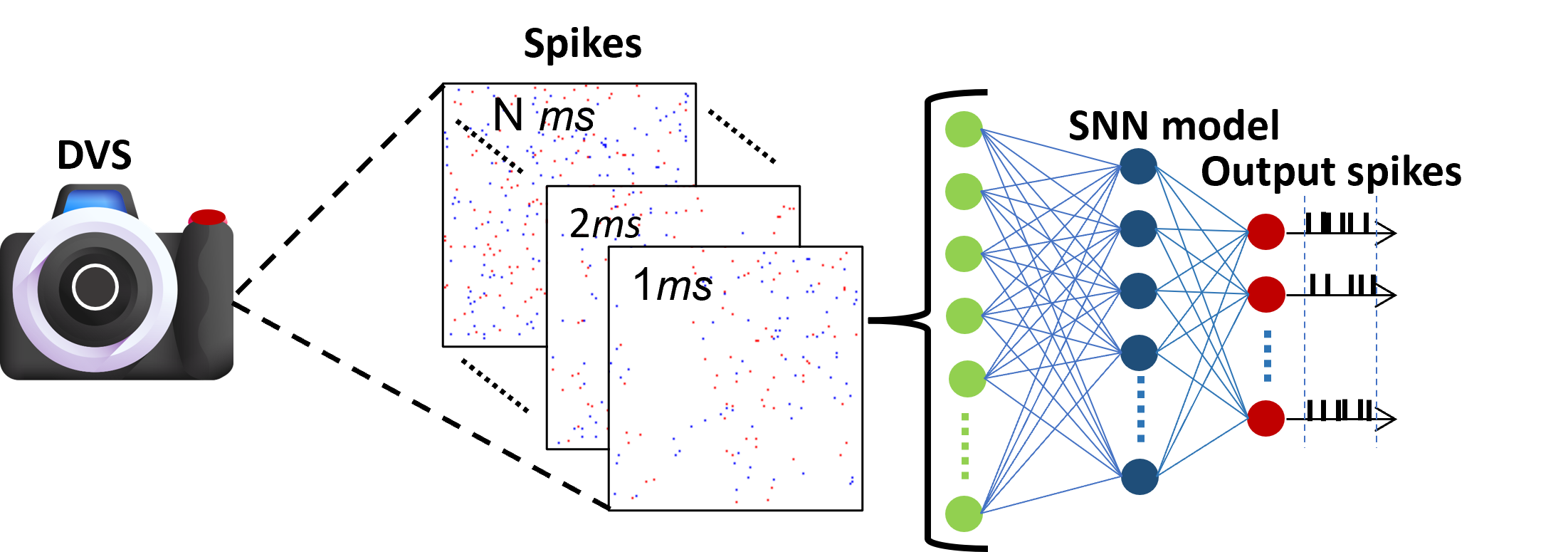}\vspace{-0.2cm}
\caption{Vision to spikes.}\vspace{-0.5cm}
\label{fig:DVS_SNN}
\end{figure}

In vision tasks \cite{OJCT15,ATBM17}, these inputs are typically generated by a DVS, which captures changes in pixel brightness rather than absolute brightness \cite{hats2018}. Each event encodes the location of a pixel, a timestamp, and a polarity indicating whether the brightness increased or decreased. This event-driven approach produces sparse and temporally precise data, making it ideal for low-power and high-speed applications. Fig. \ref{fig:DVS_SNN} shows the spike events produced from a DVS and fed to an SNN model. Red and blue dots represent the positive and negative polarity of the events accordingly.

\begin{figure}[t]
\centering
\includegraphics[width=0.92\linewidth]{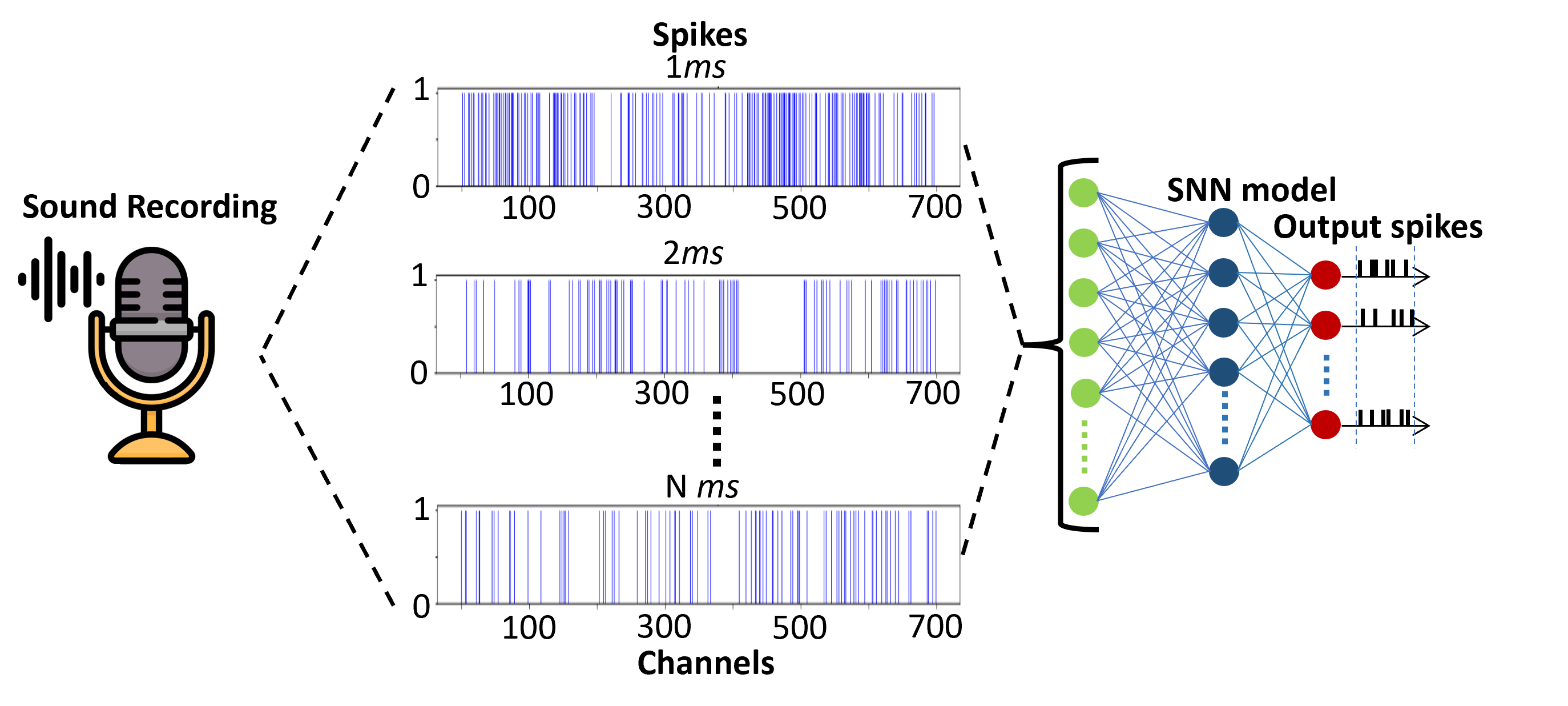}\vspace{-0.2cm}
\caption{Sound to spikes.}\vspace{-0.5cm}
\label{fig:sound_SNN}
\end{figure}

In sound tasks \cite{heidelberg}, spiking inputs are derived from recording devices that convert continuous auditory signals into spikes. This is often achieved using a bank of filters or channels that decompose the audio signal into different frequency bands, similar to the cochlea in the human ear \cite{heidelberg}. Each channel generates spikes when the energy in its frequency band exceeds a certain threshold, preserving both the temporal and spectral structure of the auditory signal. Fig. \ref{fig:sound_SNN} shows the spike events produced from a sound recording device that has $700$ channels.

These spiking representations enable SNNs to process temporal data efficiently while aligning with the asynchronous nature of neuromorphic hardware \cite{RoJaPa19, SKPMDK22}. 

\section{Proposed adversarial attacks on SNNs} \label{sec:algorithms}
As SNNs operate with binary inputs (spike or no spike), classic gradient-based techniques cannot be utilized. Hence, we designed a strategy that allows for direct gradient-based optimization exclusively in the spiking domain. The techniques used to bridge the gap between continuous gradient updates and binary spiking inputs are the Gumbel-Softmax \cite{Gumbelsoftmax2017, Gumbel_2_2017} and the Straight-Through-Estimator (STE) \cite{ste2013}. The proposed algorithms make no assumption about the architecture of the SNN, i.e., fully connected, convolutional or recurrent, and no assumption about the information coding scheme, i.e., rate coding or time-to-first-spike coding.

Input optimization in the spiking domain has also been explored in the contexts of SNN hardware accelerator testing \cite{RaSt25} and hardware Trojan attacks \cite{RKKAS25}, each using tailored real-to-binary tensor conversions and task-specific loss functions. In the first scenario, the goal is to generate a minimal-duration input sequence that maximizes fault coverage, whereas in the second, the objective is to design a trigger input capable of activating the hardware Trojan.

\subsection{Terminology and notation}

We interpret the input as being decomposed into frames of size $W \times H$ over time and we denote it by $I(t_k,x_{ij})$, where $t_k$ denotes a discrete time point and $x_{ij}$ denotes the spatial location on the frame, $i=1, \cdots, W$ and $j= 1, \cdots, H$.
If the input has duration $T$ and $T_f$ denotes the global clock period, then $k=1,\cdots,T/T_f$. At time point $t_k$, $I(t_k,x_{ij})=1$ if pixel/channel $x_{ij}$ carries a spike, otherwise $I(t_k,x_{ij})=0$ denotes no spike.

Let the SNN have $L$ layers with $N^{\ell}$ neurons in layer $\ell$, $\ell=1,\cdots,L$, with $N^{L}$ being the number of output classes.

Let $O^{\ell i}(I)$ denote the output of neuron $i$ in layer $\ell$ for input $I$ and let $O^{L}=[O^{L 1},\cdots,O^{L N^{L}}]$ denote the output of the last layer $L$. The winning class neuron of the output layer $L$ is the one that produces the largest number of spikes within a time window:

\vspace{-0.5cm}
\begin{equation}
    w = \arg\max_{i} \|O^{L i}(I)\|_1,
\end{equation}
\noindent where $\|\cdot\|_1$ is the $\ell^1$ norm and $\|O^{L w}(I)\|_1$ is the spike count of the winning class neuron $w$.
For a given dataset sample $I$, the goal of the adversarial attack is to find a slightly perturbed sample $I_{adv}$ such that a different neuron wins:
\vspace{-0.1cm}
\begin{equation}
    \arg\max_{i} \|O^{L i}(I)\|_1 \neq \arg\max_{i} \|O^{L i}(I_{adv})\|_1.
\end{equation}

Defining the variable:

\vspace{-0.1cm}
\begin{equation}
    y(k,i,j)=|I(t_k,x_{ij})-I_{adv}(t_k,x_{ij})|,\label{eq:y}
\end{equation}

\noindent the perturbation size, denoted by $PS(I,I_{adv})$, is given in \% by:\vspace{-0.1cm}
\begin{equation}
    PS(I,I_{adv})=100\cdot\frac{\sum_{k=1}^{T/T_f}\sum_{i=1}^{W}\sum_{j=1}^{H} y(k,i,j)}{W \times H \times T}.\label{eq:PS}
    \vspace{-0.1cm}
\end{equation}

\subsection{Input-specific adversarial attack}

\subsubsection{Algorithm}

\begin{figure}[t]
\centering
\includegraphics[width=0.99\linewidth]{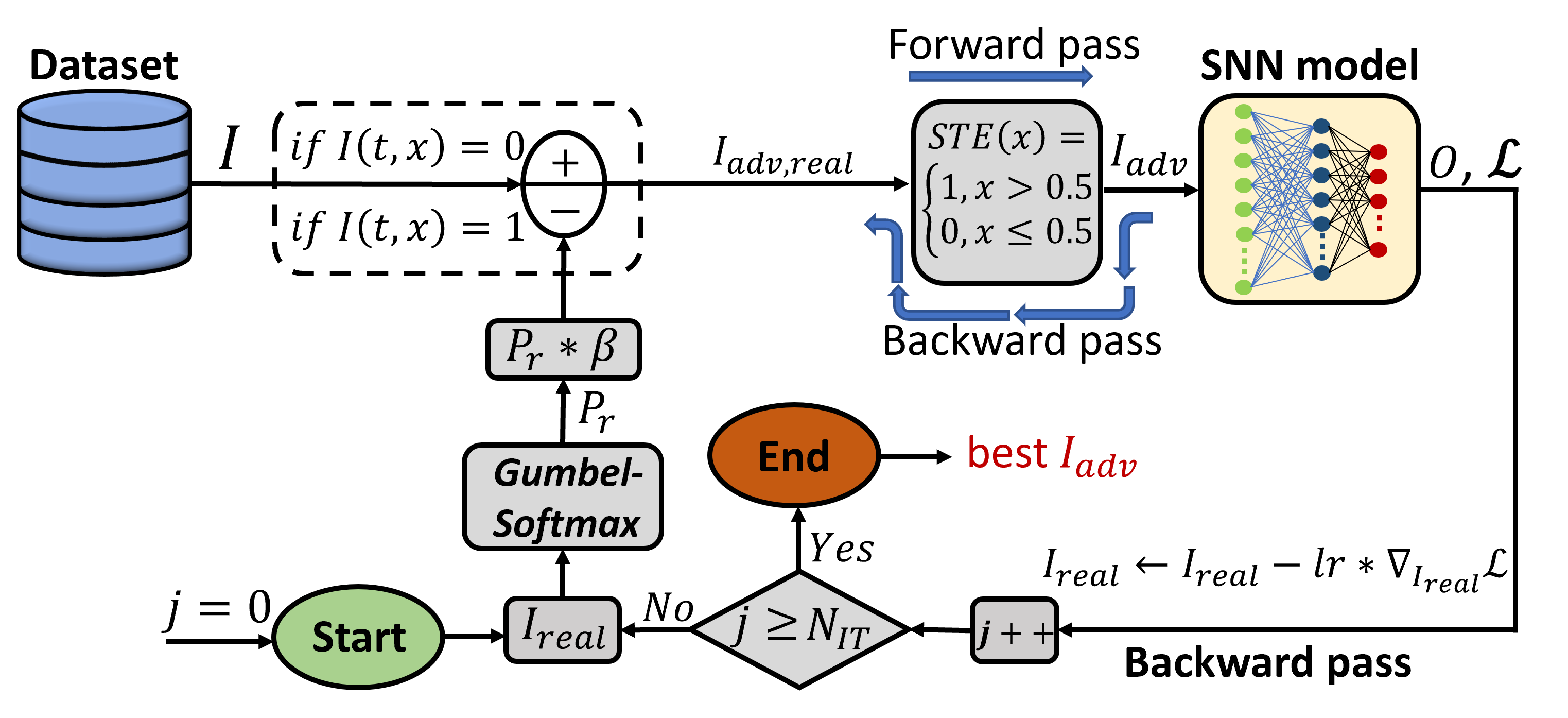}\vspace{-0.2cm}
\caption{Input-specific adversarial attack generation flow.}\vspace{-0.5cm}
\label{fig:per_sample}
\end{figure}

Fig. \ref{fig:per_sample} shows the flowchart of the input-specific adversarial attack algorithm. We start with a randomly initialized real-valued tensor $I_{real}$ of the same size as $I$. $I_{real}$ is converted into a probabilistic tensor $Pr$ with values in the range $(0,1)$ using the Gumbel-Softmax function:
\vspace{-0.1cm}
\begin{equation}
    Pr = GumbelSoftmax(I_{real}, \tau),
\end{equation}

\noindent where the temperature parameter $\tau$ controls the sharpness of the probability distribution. A high temperature produces a softer binary distribution where probabilities are closer to each other, while a lower temperature produces a sharper distribution where probabilities are closer to 0 or 1.

We generate a real-valued perturbed representation $I_{adv,real}$ of $I$ as follows:
\vspace{-0.2cm}
\begin{equation}
    I_{adv,real}(t,x) =\\ \hspace{-0.1cm} 
    \begin{cases}
        I(t,x) \hspace{-0.05cm}+\hspace{-0.05cm} Pr(t,x)\hspace{-0.05cm}*\hspace{-0.05cm}\beta & \hspace{-0.1cm} \text{if } I(t,x)\hspace{-0.05cm} = \hspace{-0.05cm}0\\
        I(t,x)\hspace{-0.05cm} -\hspace{-0.05cm} Pr(t,x)\hspace{-0.05cm}*\hspace{-0.05cm}\beta & \hspace{-0.1cm} \text{if } I(t,x)\hspace{-0.05cm} =\hspace{-0.05cm} 1
    \end{cases}
    \label{eq:perturbation_application}
\end{equation}

\noindent where $\beta$ is a scaling factor controlling the level of perturbation and is initially set to $1$.   

Next, the STE function is applied that transforms $I_{adv,real}$ using a threshold $0.5$ into the adversarial input $I_{adv}$:
\vspace{-0.1cm} 
\begin{equation}
    I_{adv} = \text{STE}(I_{adv,real}).\label{eq:I_adv}
\end{equation}

Combining Eqs. (\ref{eq:perturbation_application}) and (\ref{eq:I_adv}), we note that $Pr$ represents the probability of each event being perturbed. It is added to the elements of $I$ that are 0 and subtracted from the elements of $I$ that are 1. For $\beta=1$, the event is perturbed if $P_{r}>0.5$, i.e., a spike is added if $I(t,x)=0$ and a spike is removed if $I(t,x)=1$.

Now, the attack can be formulated as an optimization problem:
\vspace{-0.1cm}
\begin{equation}
    \vspace{-0.1cm}
    \min_{I_{real}} \mathcal{L}(I,I_{adv},O^L) \label{eq:opt_problem},
\end{equation}

\noindent where the minimization of the loss function $\mathcal{L}$ achieves the desired adversarial objective. The loss function $\mathcal{L}$ is defined as the weighted sum of three loss functions $L_i$, $i=1,\cdots,3$ that will be described in detail in Section \ref{sec:loss_functions}:
\vspace{-0.1cm} 
\begin{equation}
    \mathcal{L}= \sum_{k=1}^{3}\alpha_k * L_k,
\end{equation}

\noindent where $\alpha_i$ are the weights for scalarizing the three loss functions and aggregating them into a single one. An optimizer (e.g. Adam \cite{Adam2017optimization}) is set to apply changes to $I_{real}$ with adaptive learning rate $lr$ towards achieving the optimization objective.

Each iteration of the algorithm involves two stages. In the first stage, forward pass is performed with $I_{adv}$ and the loss function $\mathcal{L}$ is computed. In the second stage, gradient-based backpropagation is performed to compute the gradient of the loss with respect to the membrane potential of neurons in the input layer using the same backpropagation pipeline as during the training of the SNN. In our implementation, we used the SLAYER framework \cite{shor18} and did not notice the extreme gradient vanishing problem as in \cite{AdversarialG2S}. When, we reach the input layer, the STE function passes on the incoming gradient as if it was an identity function. Essentially, the STE function generates the spiking input in the forward pass while allowing the necessary continuous gradient flow during backpropagation. Thereafter, the gradients are computed normally because all tensors are real-valued and differentiable. Using the chain rule, we obtain:
\vspace{-0.1cm} 
\begin{equation}
    \nabla_{I_{real}}\mathcal{L} = \nabla_{I_{adv,real}}\mathcal{L} \cdot \frac{\partial{I_{adv,real}}}{\partial{Pr}} \cdot \frac{\partial{Pr}}{\partial{I_{real}}},
\end{equation}

\noindent where $\frac{\partial{A}}{\partial{B}}$ represents the Jacobian of the transformation from $A$ to $B$. Finally, $I_{real}$ is updated through the optimizer as follows: 
\vspace{-0.1cm} 
\begin{equation}
    I_{real}^{(l+1)} \leftarrow I_{real}^{(l)} - lr*\nabla_{I_{real}} \mathcal{L},
\end{equation}
where $I_{real}^{(l+1)}$ is the optimized $I_{real}$ after iteration $l$ of the optimization loop. The $I_{real}$ is being continuously optimized such that it leads to the ideal $P_{r}$ and eventually to the optimal $I_{adv}$. The loop terminates when a specific number of iterations $N_{IT}$ is reached and the best $I_{adv}$ that produced the minimum loss is returned, as well as the elapsed time for its generation.

\subsubsection{Loss functions}\label{sec:loss_functions}

The three loss functions are as follows:

\begin{itemize}
    \item \textit{Spatiotemporal similarity loss} $(L_1)$: Targets minimizing the spatiotemporal similarity between the original and perturbed samples. For this purpose, it uses the sample variance of the variable $y(k,i,j)$ in Eq. (\ref{eq:y}) denoted by $\mbox{Var}(y)$:
    \begin{equation}
    L_1 = max(0,\mbox{Var}(y)-r_1).
        \label{eq:L1}
    \end{equation}
    Relaxation parameter $r_1$ is adaptive starting from $0$ and increasing every few iterations if there is no improvement in the loss. We use the sample variance instead of the perturbation size in Eq. (\ref{eq:PS}) as we experimentally found that the optimization converges faster.
     
    \item \textit{Winning class loss} $(L_2)$: Let $w$ denote the winning class neuron when forward passing sample $I$, i.e., $w = \arg\max_{i} \|O^{L i}(I)\|_1$. This loss function measures the spike counts of this winning neuron when forward passing the adversarial sample $I_{adv}$:
    \vspace{-0.1cm}
    \begin{equation}
        L_2 = \|O^{L w}(I_{adv})\|_1.
        \label{eq:L2}
        \vspace{-0.1cm}
    \end{equation}
    The idea behind this loss function lies in the fact that if the spikes of the originally winning class are gradually reduced, 
    then eventually the winning class will change achieving the adversarial objective.
    \item \textit{Confidence margin loss} $(L_3)$: At every iteration, $I_{adv}$ is optimized to force the network into misclassifying it.
    When the winning class changes from $w$ to $w_{adv}$, this loss function ensures that there is a minimum spike count difference $d$ between the original and new winning classes: 
    \vspace{-0.15cm}
    \begin{equation}
        L_3 \hspace{-0.05cm}= \hspace{-0.05cm}max(0,\hspace{-0.055cm}\|O^{L w}(\hspace{-0.05cm}I_{adv}\hspace{-0.05cm})\|_1 \hspace{-0.07cm}-\hspace{-0.05cm} \|O^{L w_{adv}}(\hspace{-0.05cm}I_{adv}\hspace{-0.05cm})\|_1  \hspace{-0.05cm}+\hspace{-0.05cm} d).
        \vspace{-0.1cm}
    \end{equation}
\end{itemize}

The direct gradient application and adaptive hyper-parameters (e.g., $lr$, $\tau$, $\beta$, $r_1$, $d$) make the optimization to converge quickly and always lead to the generation of an adversarial example. 

\subsection{Universal Adversarial Attack}

\begin{figure}[t]
\centering
\includegraphics[width=0.99\linewidth]{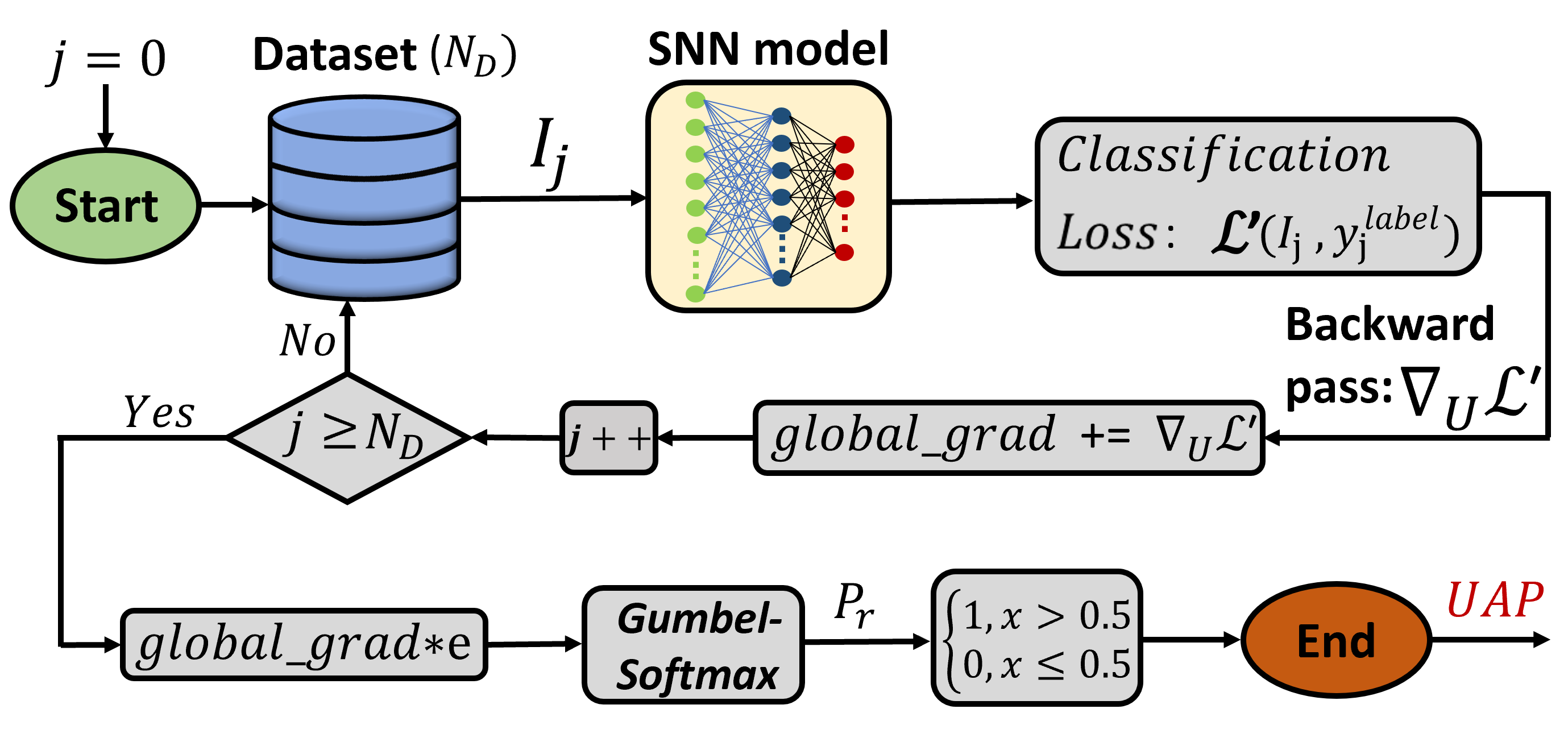}\vspace{-0.2cm}
\caption{Universal adversarial attack generation flow.}\vspace{-0.2cm}
\label{fig:upa_attack}
\end{figure}

Fig. \ref{fig:upa_attack} shows the flowchart of the proposed universal adversarial attack generation algorithm. The steps are as follows:

\begin{enumerate}
    \item The dataset of size $N_D$ used to train and validate the SNN model is being loaded and the model is set again to training mode.
    \item A global tensor $global\_grad$ is defined having the same size as input $I$. $global\_grad$ is initialized with zeros:
    \vspace{-0.1cm}
    \begin{equation}
        global\_grad = 0.
        \vspace{-0.1cm}
    \end{equation}
    The goal is to progressively update this tensor with the gradient information gathered from the dataset samples.
    
    \item Every sample $I_j$ from the dataset is forward passed through the network in order to calculate the classification loss $\mathcal{L}^{'}(I_j,y_j^{label})$, where $y_j^{label}$ is the target class of $I_j$.
    \item Backpropagation is performed to obtain the gradients of the loss with respect to the membrane potential of the neurons of the first layer, denoted by $U$. These gradients, denoted by $\nabla_U\mathcal{L}^{'}$, indicate the direction and the magnitude in which the membrane potential of the neurons should change to either increase or decrease the likelihood of a correct classification. By extension, the membrane potential is defined by the number and rate of incoming spikes, thus $\nabla_U\mathcal{L}^{'}$ indirectly indicates the change in the input spike train.
    \item During the inference, the gradients are accumulated in the global gradient tensor $global\_grad$: 
    \vspace{-0.07cm}
    \begin{equation}
        global\_grad += \nabla_U\mathcal{L}^{'}.
        \vspace{-0.07cm}
    \end{equation}
    The purpose of accumulating the gradients is to capture global patterns of sensitivity. Certain input neurons will consistently show higher gradient magnitude across multiple samples, which suggests that they are critical to the decision making. These critical input neurons are the primary target for perturbing their incoming spike trains.
    \item After the inference over the complete dataset is finished,
    the Gumbel-Softmax function is applied to convert the real-value global gradient tensor into the range $(0,1)$:
    \vspace{-0.05cm}
    \begin{equation}
        P_{r} = GumbelSoftmax(global\_grad \times e, \tau),
        \vspace{-0.1cm}
    \end{equation}
    where $e$ controls the perturbation magnitude and initially is set to $1$. $Pr$ here represents the probabilities of input spikes being perturbed. Values closer to $1$ mean that the corresponding spike is highly likely to be perturbed, whereas values close to $0$ indicate that the corresponding spike most likely will remain unchanged. $\tau$ is crucial as it controls the sharpness of the perturbations. Higher $\tau$ leads to smoother perturbations whereas lower $\tau$ leads to more sharp perturbations. 
    \item In order to apply $Pr$ to a sample of the dataset, it needs to be in the binary domain. A rounding function gives the final universal adversarial patch $UAP$ as follows:
    \begin{equation}
        UAP = round(Pr).
    \end{equation}
\end{enumerate}

The stealthiness of the UAP is defined based on the spatiotemporal sparsity of its spikes. It becomes a perturbation when added to the incoming input. The perturbation can be expressed using the same metric in Eq. \ref{eq:PS} by setting $I=\boldsymbol{0}$. A single dataset inference suffices to generate the $UAP$, thus the adversarial example generation time is fixed and equals the dataset inference time. A second inference is performed to validate the ASR of the $UAP$.

\begin{figure}[t]
\centering
\includegraphics[width=0.94\linewidth]{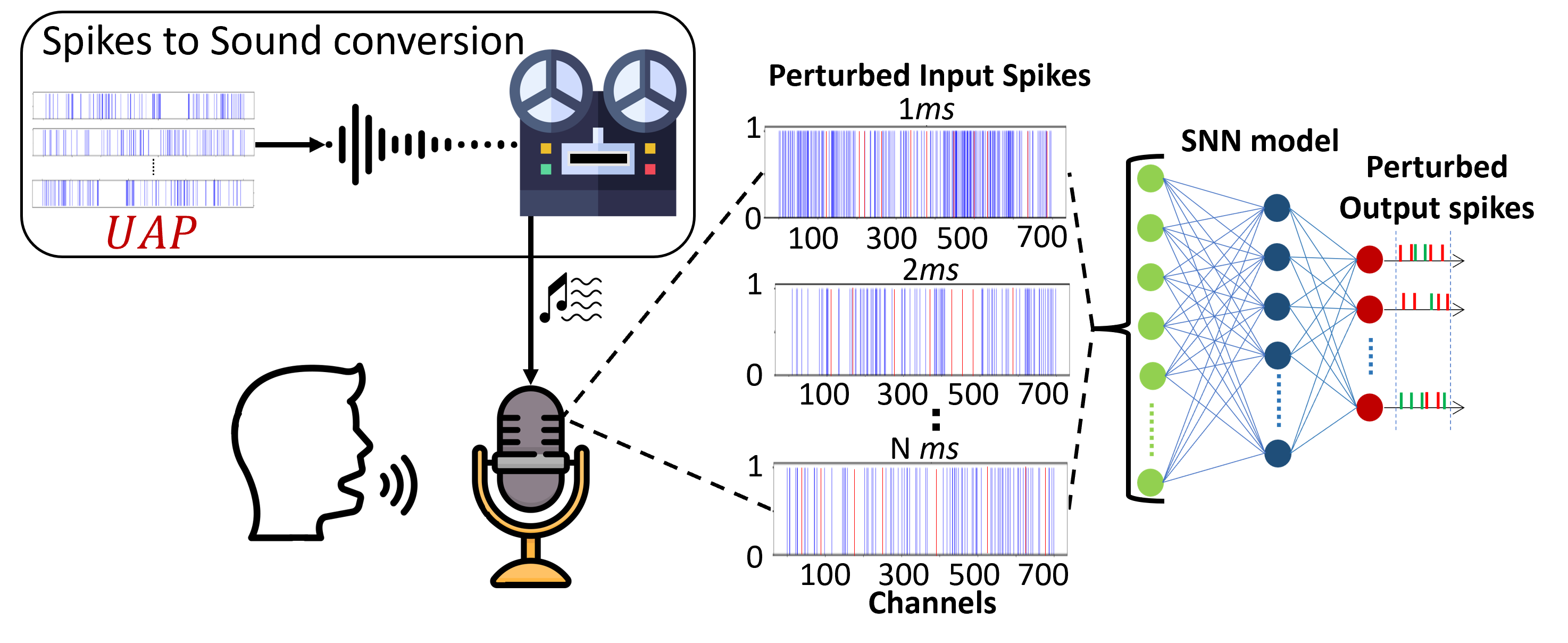}\vspace{-0.1cm}
\caption{Example universal adversarial attack application in the sound domain.}\vspace{-0.75cm}
\label{fig:real_case_attack_sound}
\end{figure}

\begin{figure}[t]
\centering
\includegraphics[width=0.71\linewidth]{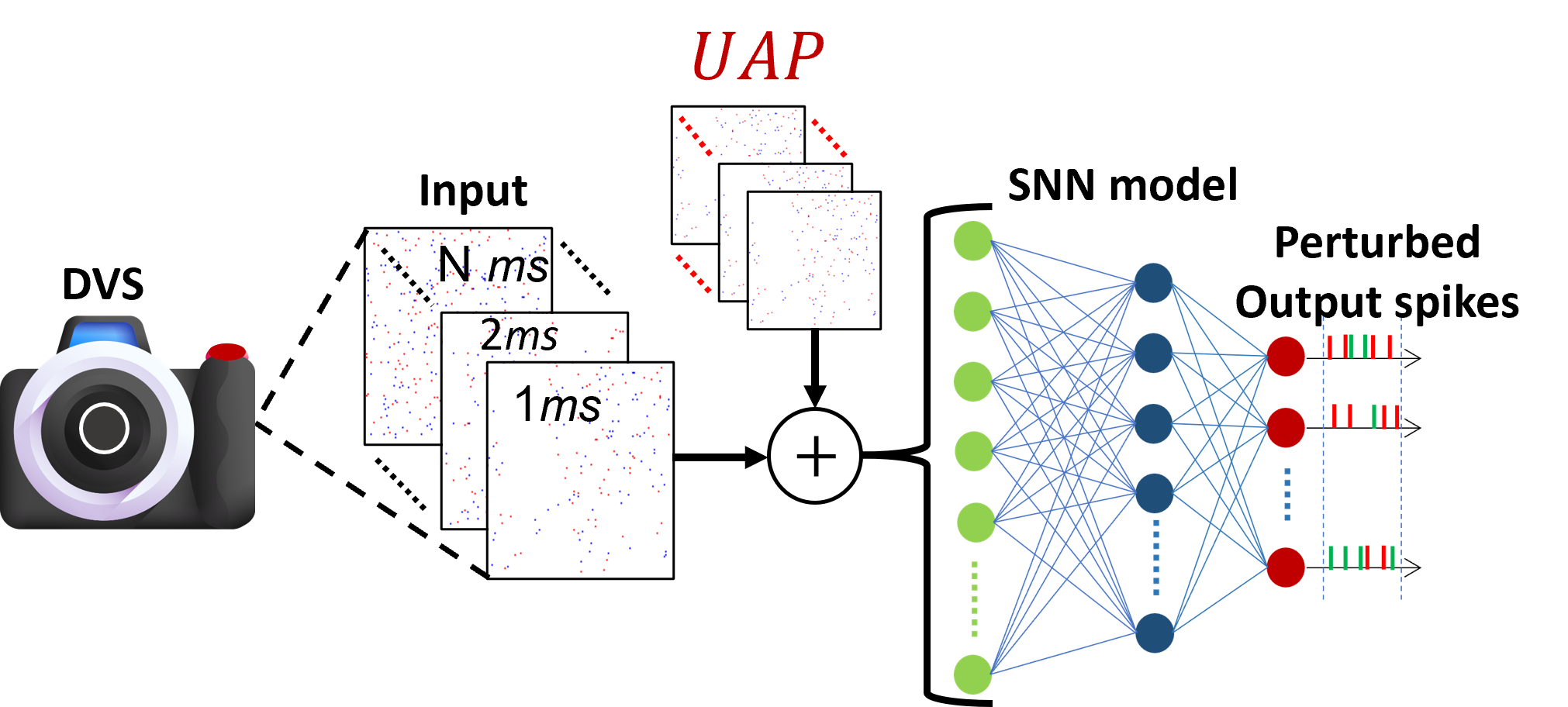}\vspace{-0.1cm}
\caption{Example universal adversarial attack application in the vision domain.}\vspace{-0.4cm}
\label{fig:real_case_attack_image}
\end{figure}

The final $UAP$ can be added to any input $I$ during the deployment of the SNN model to fool the prediction. In the sound domain, the spike-based information from the $UAP$ can be transformed into an audio signal by mapping the spike events to corresponding audio frequencies. This perturbation can then be recorded and replayed to a sound-to-spike sensor, causing the target SNN to misinterpret the input. Fig. \ref{fig:real_case_attack_sound} illustrates this attack flow, where human voice commands are perturbed by the $UAP$, leading to confusion in the SNN. In the vision domain, the spike-based information from the $UAP$ can be injected into the event stream generated by a DVS. This modification alters the spatiotemporal event patterns before they reach the SNN, resulting in incorrect model outputs. Fig. \ref{fig:real_case_attack_image} illustrates this attack flow. 

\section{Experimental setup} \label{sec:exp_setup}

\begin{figure}[t!]
\centering
\includegraphics[width=1\linewidth]{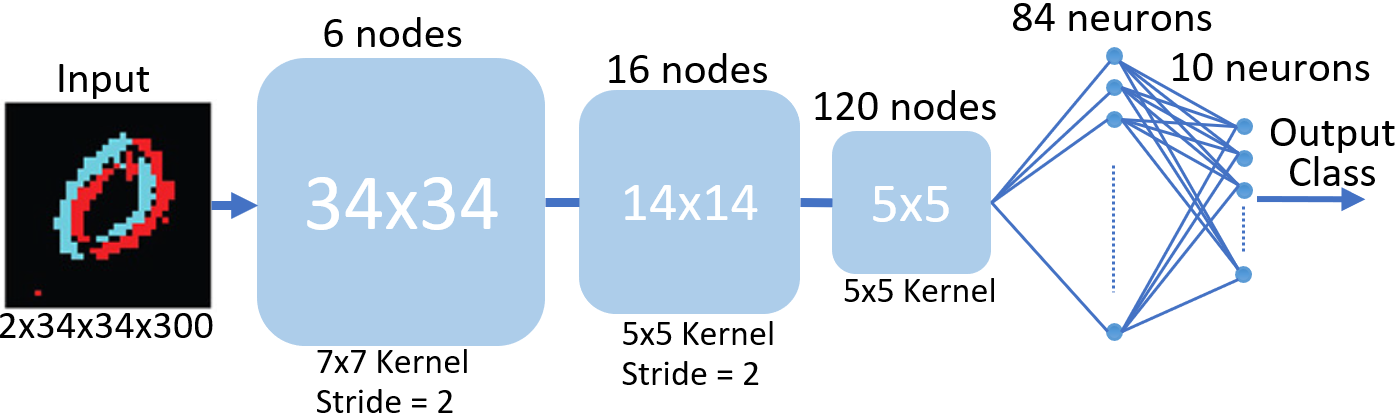}\vspace{-0.1cm}
\caption{SNN architecture for the NMNIST dataset.}\vspace{-0.4cm}
\label{fig:nmnist_SNN}
\end{figure}

\begin{figure}[t]
\centering
\includegraphics[width=1.0\linewidth]{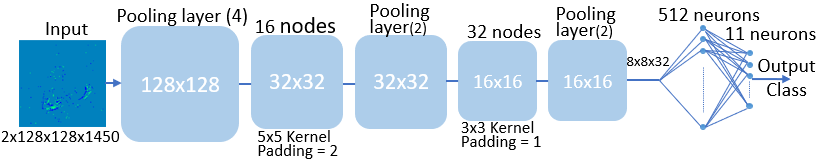}\vspace{-0.1cm}
\caption{SNN architecture for the IBM DVS Gesture dataset.}\vspace{-0.4cm}
\label{fig:IBM_SNN}
\end{figure}

\begin{figure}[t]
\centering
\includegraphics[width=0.9\linewidth]{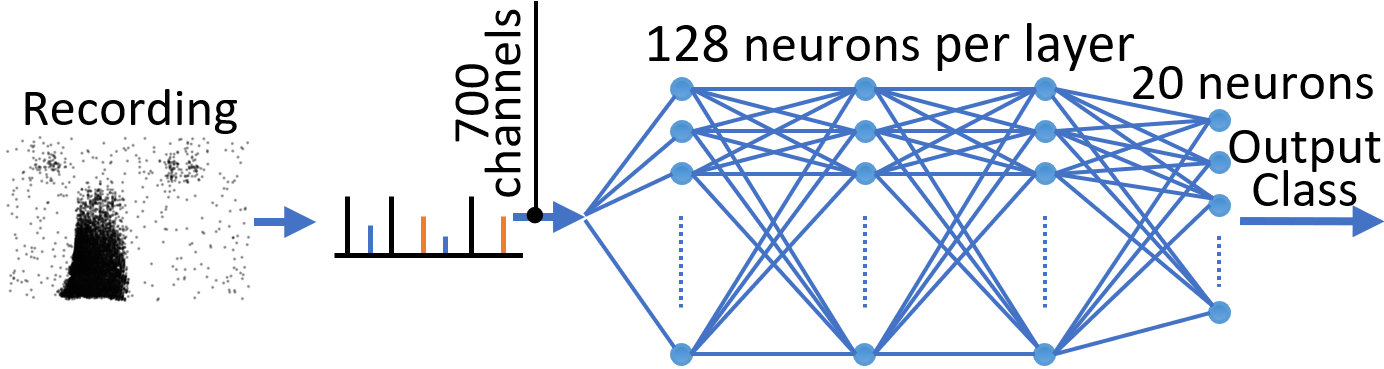}\vspace{-0.1cm}
\caption{SNN architecture for the SHD dataset.}
\label{fig:SHD_SNN}
\vspace{-0.0cm}
\end{figure}

The adversarial attack algorithms are evaluated on three datasets, namely NMNIST, IBM DVS Gesture, and SHD. The SNN training and gradient-based backpropagation for attack generation were performed using the SLAYER framework \cite{shor18}. NMNIST is a spiking version of MNIST that consists of 60K training and 10K testing images of handwritten digits, captured by a DVS while it views MNIST images on an LCD monitor \cite{OJCT15}. The IBM DVS Gesture dataset contains 1341 samples of 11 hand and arm gestures performed by 29 individuals under three lighting conditions, and is also recorded via a DVS \cite{ATBM17}. The SHD dataset includes 8332 training and 2088 testing samples of spoken English and German digits, converted into spike trains using a neuromorphic cochlea model, spanning in total 20 classes \cite{heidelberg}. The corresponding SNN architectures are shown in Figs. \ref{fig:nmnist_SNN}-\ref{fig:SHD_SNN}. Table \ref{tab:SNN_characteristics} summarizes the SNN characteristics, including the nominal prediction accuracy, the input spatial and temporal dimensions, and the input spikes size considering an 1$ms$ timestep.

\begin{table}[t]
\centering
\scriptsize
\caption{Benchmark SNNs characteristics.}
\vspace{-0.15cm}
\begin{tabular}{ | m{2.6cm} | m{1.2cm}| m{1.5cm} | m{1.2cm} | } 
  \hline
   & \textbf{NMNIST} & \textbf{IBM} & \textbf{SHD}\\
  \hline
  Prediction accuracy & $98.19\%$ & $86.36\%$ & $76.59\%$\\  
  \hline
  \# Output classes & $10$ & $11$ & $20$\\ 
  \hline
  Input spatial dimension & $2 \times 34 \times 34$ & $2\times128\times128$ & $700 \times 1 \times 1$\\ 
  \hline
  Input temporal dimension & $300 \, ms$ & $1.45 \, s$ & $~1 \, s$\\ 
  \hline
  Input spikes size & $693600$ & $47513600$ & $700000$\\ 
  \hline
  Size training set & $60K$ & $1080$ & $8332$\\ 
  \hline
  Size testing set & $10K$ & $261$ & $2088$ \\ 
  \hline
\end{tabular}
\vspace{-0.4cm}
\label{tab:SNN_characteristics}
\end{table}
\vspace{-0.1cm}
\section{Results}\label{sec:results}

\subsection{Input-specific adversarial attack}

\begin{table}[t]
\centering
\scriptsize
\caption{Input-specific adversarial attack results.}
\vspace{-0.15cm}
\begin{tabular}{ | m{3.0cm} | m{1.1cm}| m{1.1cm} | m{1.1cm} | } 
  \hline
   & \textbf{NMNIST} & \textbf{IBM} & \textbf{SHD}\\
  \hline
  Samples tested & $69369$ & $1304$ & $9921$\\
  \hline
  ASR& $100\%$ & $100\%$ & $100\%$\\ 
  \hline
  Average perturbation & $0.0305\%$ & $0.0018\%$ & $0.0585\%$\\
  \hline
  Minimum generation time& $0.0121s$ & $0.122s$ & $0.0061s$\\
  \hline
  Average generation time& $2.277 s$ & $2.465s$ & $9.7688s$\\
  \hline
  Maximum generation time& $6.3085s$ & $4.682s$ & $36.4226s$\\ 
  \hline
  Average number of iterations& $114$ & $74$ & $1008$\\ 
  \hline
\end{tabular}
\vspace{-0.2cm}
\label{tab:per_sample_results}
\end{table}

\begin{figure}[h!]
\begin{subfigure}{0.24\textwidth}
  \centering
  \includegraphics[width=0.78\linewidth]{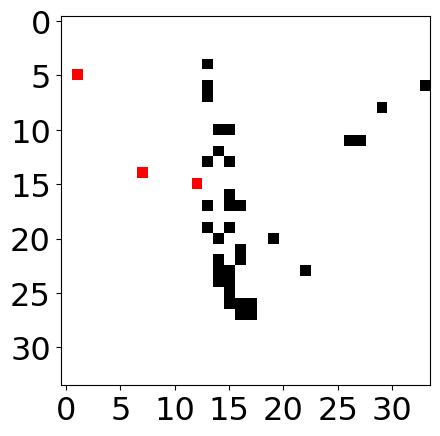}
  \vspace{-0.2cm}
  \caption*{30-40ms}
  \label{fig:sfig1}
\end{subfigure}%
\begin{subfigure}{.24\textwidth}
  \centering
  \includegraphics[width=.78\linewidth]{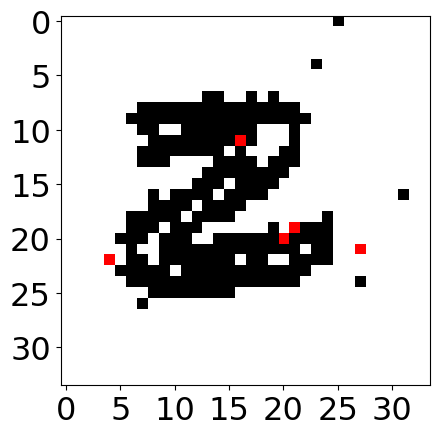}
  \vspace{-0.2cm}
  \caption*{30-40ms}
  \label{fig:sfig2}
\end{subfigure}
\begin{subfigure}{0.24\textwidth}
  \centering
  \includegraphics[width=0.78\linewidth]{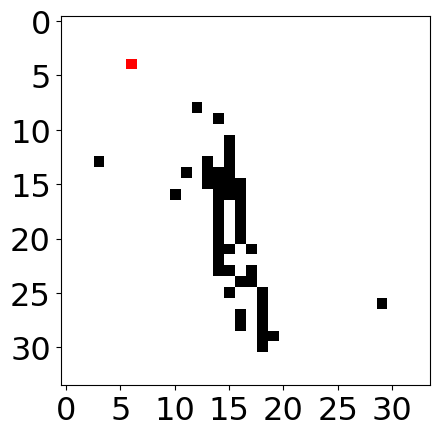}
  \vspace{-0.2cm}
  \caption*{70-80ms}
  \label{fig:sfig3}
\end{subfigure}%
\begin{subfigure}{.24\textwidth}
  \centering
  \includegraphics[width=.78\linewidth]{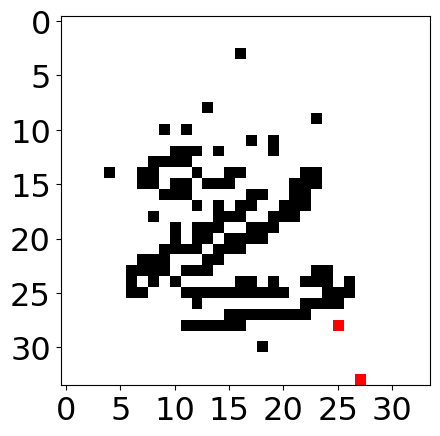}
  \vspace{-0.2cm}
  \caption*{70-80ms}
  \label{fig:sfig4}
\end{subfigure}
\begin{subfigure}{.24\textwidth}
  \centering
  \includegraphics[width=.78\linewidth]{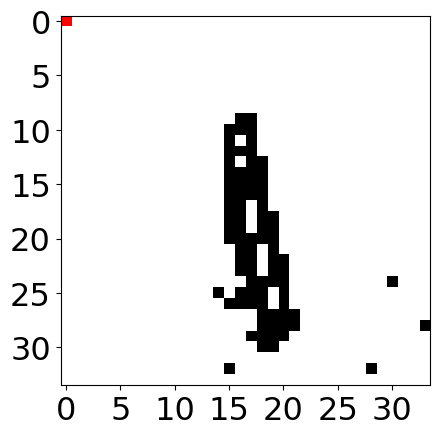}
  \vspace{-0.2cm}
  \caption*{130-140ms}
  \label{fig:sfig5}
\end{subfigure}
\begin{subfigure}{.24\textwidth}
  \centering
  \includegraphics[width=.78\linewidth]{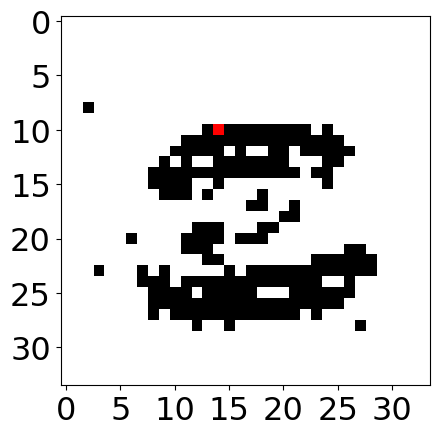}
  \vspace{-0.2cm}
  \caption*{130-140ms}
  \label{fig:sfig6}
\end{subfigure}
\begin{subfigure}{.24\textwidth}
  \centering
  \includegraphics[width=.9\linewidth]{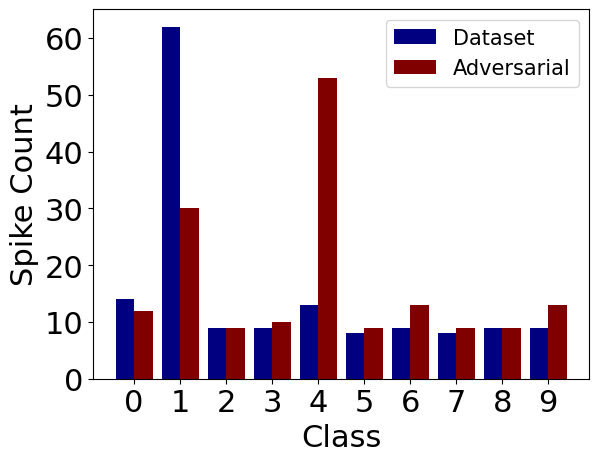}
  \vspace{-0.2cm}
  \caption{1 classified as 4.}
  \label{fig:sfig7}
\end{subfigure}
\begin{subfigure}{.24\textwidth}
  \centering
  \includegraphics[width=.9\linewidth]{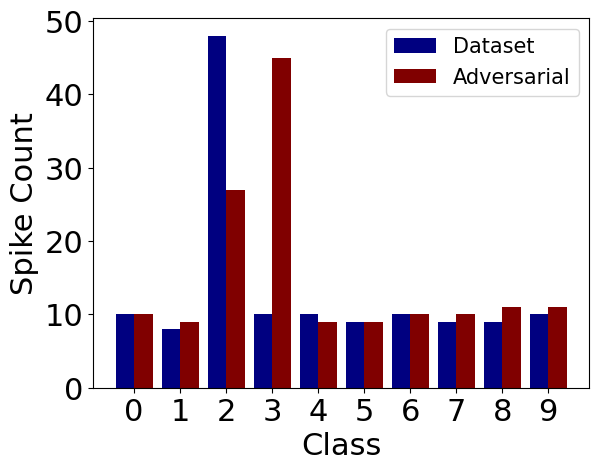}
  \vspace{-0.2cm}
  \caption{2 classified as 3.}
  \label{fig:sfig8}
\end{subfigure}
\caption{Adversarial examples for the N-MNIST dataset.}
\label{fig:nmnist_adversarial}
\vspace{-0.4cm}
\end{figure}

\begin{figure}[h!]
\begin{subfigure}{0.24\textwidth}
  \centering
  \includegraphics[width=0.78\linewidth]{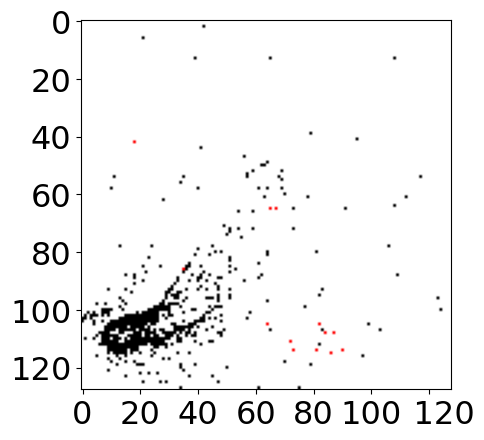}
  \vspace{-0.2cm}
  \caption*{100-110ms}
  \label{fig:sfig1}
\end{subfigure}%
\begin{subfigure}{.24\textwidth}
  \centering
  \includegraphics[width=.78\linewidth]{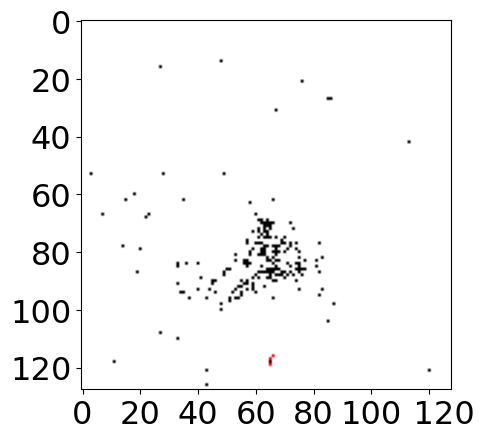}
  \vspace{-0.2cm}
  \caption*{100-110ms}
  \label{fig:sfig2}
\end{subfigure}
\begin{subfigure}{0.24\textwidth}
  \centering
  \includegraphics[width=0.78\linewidth]{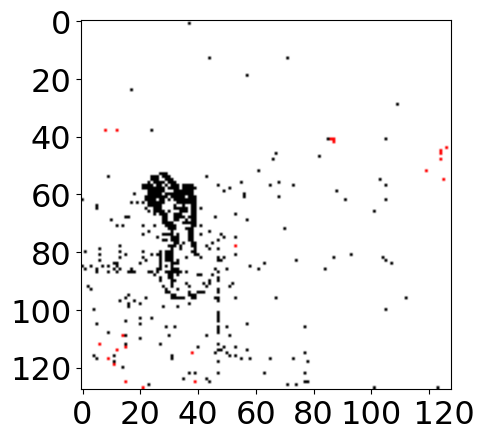}
  \vspace{-0.2cm}
  \caption*{500-510ms}
  \label{fig:sfig3}
\end{subfigure}%
\begin{subfigure}{.24\textwidth}
  \centering
  \includegraphics[width=.78\linewidth]{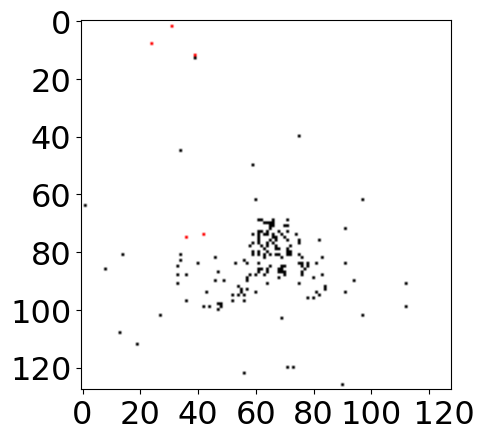}
  \vspace{-0.2cm}
  \caption*{500-510ms}
  \label{fig:sfig4}
\end{subfigure}
\begin{subfigure}{.24\textwidth}
  \centering
  \includegraphics[width=.78\linewidth]{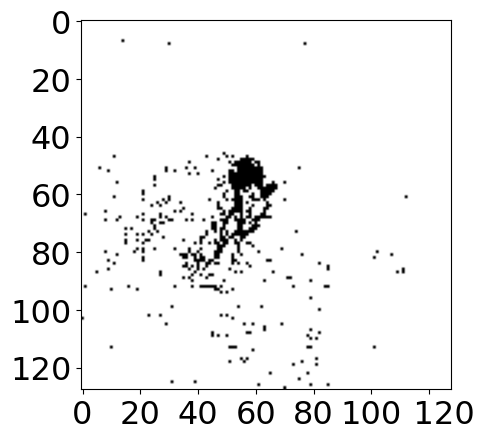}
  \vspace{-0.2cm}
  \caption*{900-910ms}
  \label{fig:sfig4}
\end{subfigure}
\begin{subfigure}{.24\textwidth}
  \centering
  \includegraphics[width=.78\linewidth]{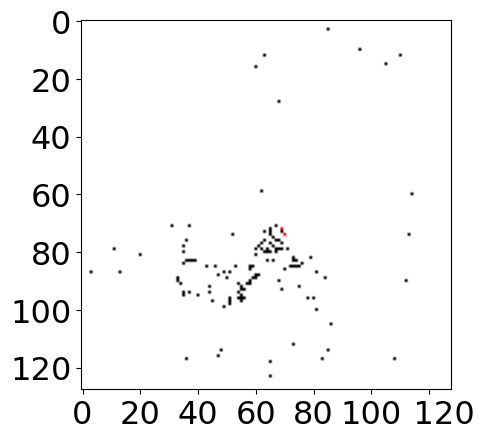}
  \vspace{-0.2cm}
  \caption*{900-910ms}
  \label{fig:sfig4}
\end{subfigure}
\begin{subfigure}{.24\textwidth}
  \centering
  \includegraphics[width=.9\linewidth]{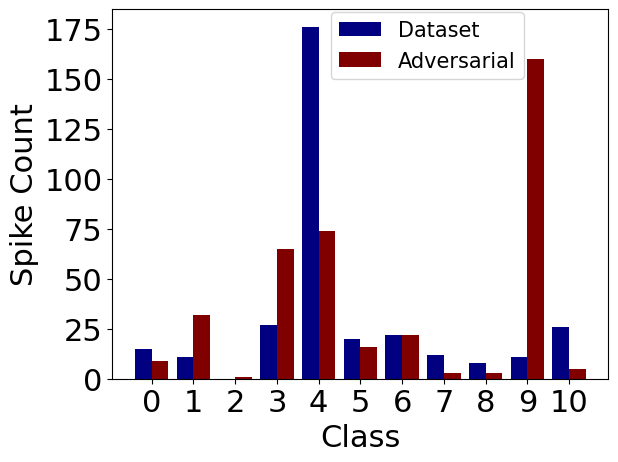}
  \vspace{-0.2cm}
  \caption{Right arm counter clockwise classified as playing guitar.}
  \label{fig:sfig4}
\end{subfigure}
\begin{subfigure}{.24\textwidth}
  \centering
  \includegraphics[width=.9\linewidth]{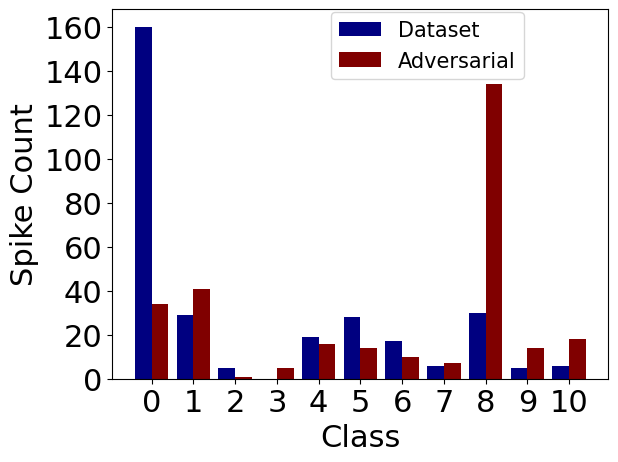}
  \vspace{-0.2cm}
  \caption{Hand clapping classified as playing drums.}
  \label{fig:sfig4}
\end{subfigure}
\caption{Adversarial examples for the IBM DVS Gesture dataset.} 
\label{fig:IBM_adversarial}
\vspace{-0.2cm}
\end{figure}

Table \ref{tab:per_sample_results} summarizes the results of the input-specific adversarial attack for the three case studies. We evaluated only the samples that were correctly classified, as generating an adversarial example for a misclassified sample would be meaningless. The algorithm successfully generates an adversarial example for all tested samples, achieving an ASR of $100\%$. The maximum average perturbation across the three case studies is 0.0585\%, rendering the attacks extremely stealthy. Table \ref{tab:per_sample_results} also shows the minimum, maximum, and average adversarial example generation time, as well as the average number of iterations across the tested samples. Figs. \ref{fig:nmnist_adversarial} and \ref{fig:IBM_adversarial} illustrate for the NMNIST and IBM DVS Gesture case studies, respectively, the result for two tested samples. Each column corresponds to one tested sample. The first three sub-plots show three snapshots of the adversarial input, where, for the purpose of better visualization, each snapshot projects the spikes of 10 consecutive timesteps onto one plane. Black and red dots indicate original and perturbed spikes, respectively. The subplot at the bottom of the column shows the spike count of output neurons corresponding to the different classes for the original dataset sample and its adversarial example. The results demonstrate that minor perturbations in spike trains, forming an adversarial example nearly indistinguishable from the original sample, consistently caused misclassification. This underscores both the effectiveness of the proposed algorithm in terms of ASR and strealthiness, and the susceptibility of SNNs to the introduced adversarial attack.

\subsection{Universal adversarial attack}

\begin{table}[t]
\centering
\scriptsize
\caption{Universal adversarial attack results.}
\vspace{-0.15cm}
\begin{tabular}{ | m{2.2cm} | m{1.2cm}| m{1.5cm} | m{1.2cm} | } 
  \hline
   & \textbf{NMNIST} & \textbf{IBM} & \textbf{SHD}\\
  \hline
  Samples tested & $69369$ & $1304$ & $9921$\\
  \hline
  ASR& $81.66\%$ & $87.66\%$ & $78.35\%$\\
 
  \hline
  Average perturbation& $0.45\%$ & $0.079\%$ & $0.57\%$\\ 
  \hline
\end{tabular}
\label{tab:uap_results}
\vspace{-0.4cm}
\end{table}

\begin{figure}[t]
\centering
\begin{subfigure}[b]{0.2\textwidth}
  \centering
  \includegraphics[width=0.82\textwidth]{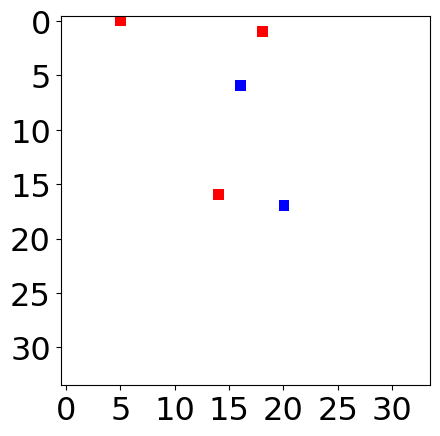}
  \vspace{-0.1cm}
  \caption{NMNIST (10ms).}
  \label{fig:sfig1}
\end{subfigure}%
\hspace{0.01\textwidth}
\begin{subfigure}[b]{.2\textwidth}
  \centering
  \includegraphics[width=0.9\textwidth]{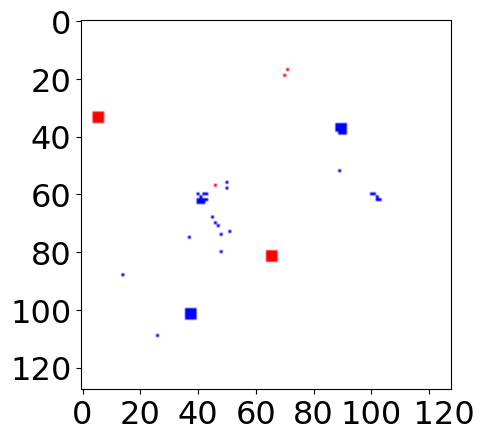}
  \vspace{-0.1cm}
  \caption{IBM DVS Gesture (10ms).}
  \label{fig:sfig2}
\end{subfigure}
\begin{subfigure}[b]{0.35\textwidth}
  \centering
  \includegraphics[width=\textwidth]{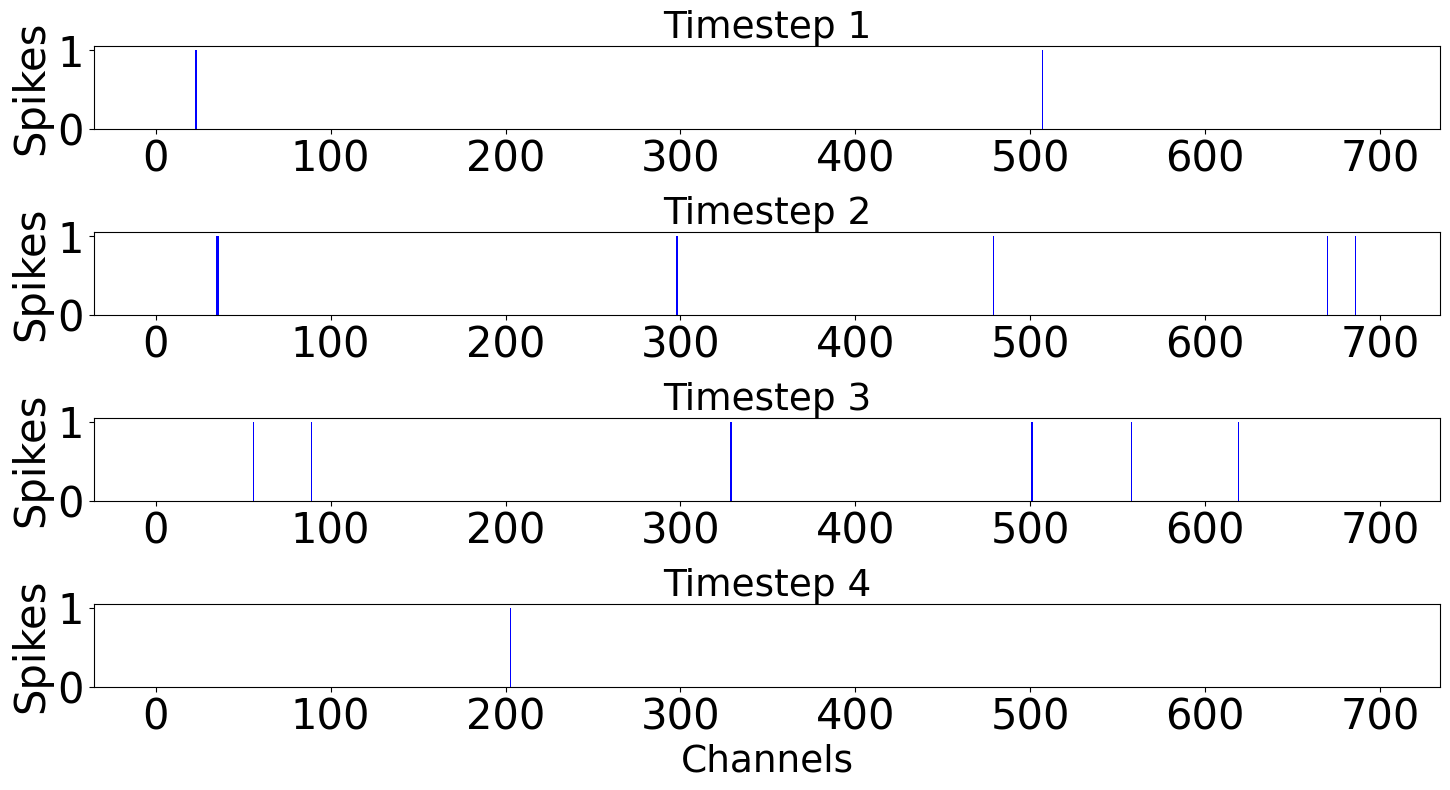}
  \vspace{-0.4cm}
  \caption{SHD (4ms).}
  \label{fig:sfig3}
\end{subfigure}%
\vspace{-0.1cm}
\caption{$UAP$ illustration.}
\vspace{-0.4cm}
\label{fig:UAP_spike_image}
\end{figure}

Table \ref{tab:uap_results} presents the performance of the proposed universal adversarial attack across the three case studies. As it can be seen, mixing inputs with the $UAP$ results in low perturbation and consistently high ASR. Fig. \ref{fig:UAP_spike_image} visualizes the spike pattern of the $UAP$ for each case study. For the MNIST and IBM we show three snapshots that accumulate the spikes across 10 timesteps, while for the SHD we show the spike pattern across the channels for the first 4 timesteps. The $UAP$ shows a sparse spike pattern and when added to the incoming sample becomes imperceptible. Although the perturbation is larger compared to the input-specific adversarial attack, the advantage of the $UAP$ is that it can manipulate the vast majority of inputs in real-time, fooling the network's decision.

\subsection{Comparison with the state-of-the-art}

\subsubsection{Input-specific adversarial attack}

\begin{table}[t]
\centering
\scriptsize
\caption{Comparison of input-specific adversarial attacks on the NMNIST.}
\vspace{-0.15cm}
\begin{tabular}{ | m{1cm} | m{0.7cm} |m{0.7cm} | m{0.9cm} | m{1.4cm} | m{1.2cm} |} 
  \hline
   \textbf{Attack} & \textbf{Samples tested} & \textbf{ASR} & \textbf{Perturba-tion} & \textbf{Average generation time/frame} & \textbf{Input spikes size}\\
  \hline
    \cite{marchisio2021dvsattacksadversarialattacksdynamic} & 1000 & $74.44\%$ & \centering$23.28\%$ & $0.0106\, s$  & $138720$\\ 
  \hline
  \cite{spikeFool} $(1)$  & 1000 & $99.53\%$ & $0.1845\%$ & $0.4145\, s$ & $138720$\\
  \hline
  \cite{spikeFool} $(2)$  & 1000 & $99.88\%$ & $0.196\%$ & $0.2\, s$  & $138720$\\
  \hline 
  \cite{AdversarialG2S} & 1000 & $100\%$ & $0.6237\%$ & $0.015\, s$  & $138720$\\
  \hline
  This work & $69369$ & $100\%$ & $0.0305\%$ & $0.00759\, s$  & $693600$\\ 
  \hline
\end{tabular}
\label{tab:comparison_1}
\vspace{-0.2cm}
\end{table}

\begin{table}[t]
\centering
\scriptsize
\caption{Comparison of input-specific adversarial attacks on the IBM DVS Gesture.}
\vspace{-0.15cm}
\begin{tabular}{ | m{1cm} | m{0.7cm} | m{0.7cm} | m{0.9cm} | m{1.4cm} |m{1.2cm} |} 
  \hline
   \textbf{Attack} & \textbf{Samples tested} & \textbf{ASR} & \textbf{Perturba-tion} & \textbf{Average generation time/frame} & \textbf{Input spikes size}\\
  \hline
    \cite{marchisio2021dvsattacksadversarialattacksdynamic} & 1000 & $92.44\%$ & \centering$4.78\%$ & $9.65e^{-4}\, s$ & $4751360$\\ 
  \hline
  \cite{spikeFool} $(1)$ & 1000 & $100\%$ & $0.0065\%$ & $0.0176\, s$ & $4751360$\\
  \hline
  \cite{spikeFool} $(2)$ & 1000  & $99.87\%$ & $0.0042\%$ & $0.0165\, s$ & $4751360$\\
  \hline
  \cite{spikeFool} $(3)$ & 1000  & $97.61\%$ & $0.0024\%$ & $0.0196\, s$& $4751360$\\
  \hline 
  \cite{AdversarialG2S} & 1000& $99.77\%$ & $0.028\%$ & $0.0036\, s$ & $4751360$\\
  \hline
  This work & $1304$ & $100\%$ & $0.0018\%$ & $0.0017\, s$& $47513600$\\ 
  \hline
\end{tabular}
\label{tab:comparison_2}
\vspace{-0.4cm}
\end{table}

Tables \ref{tab:comparison_1} and \ref{tab:comparison_2} present for the NMNIST and IBM DVS Gesture case studies, respectively, a comparative analysis with the respect to the state-of-the-art gradient-based attacks, namely the \textit{sparse} attack \cite{marchisio2021dvsattacksadversarialattacksdynamic}, the \textit{SpikeFool} attack \cite{spikeFool}, and the attack in \cite{AdversarialG2S}. For the \textit{SpikeFool} attack, more than one result is shown exploring the trade-off between ASR and perturbation. The evaluation metrics for these prior attacks were taken from \cite{spikeFool}. Therein, the spikes of each recording were binned into a reduced set of timesteps (e.g. 60 for the NMNIST and 145 for the IBM DVS Gesture), capping the maximum number of spikes to 1 per pixel. This reduced the input spikes size significantly from $2\times34\times34\times300=693600$ to $2\times34\times34\times60=138720$ for the NMNIST and from $2\times128\times128\times1450=47513600$ to $2\times128\times128\times145=4751360$ for the IBM DVS Gesture. Moreover, only 1000 samples were tested. In contrast, the results for our proposed method are produced considering the full input spike size and all samples in the dataset that are correctly classified. In other words, the ASR for the prior attacks is likely to be optimistic as it is computed on a sub-dataset and, in addition, our algorithm was executed searching in the much larger raw input spikes space, thus solving a more challenging optimization. Still, the results show that are our attack outperforms all prior attacks in all metrics.

\subsubsection{Universal adversarial attack}

\begin{table}[t]
\centering
\scriptsize
\caption{Comparison of universal adversarial attacks on the IBM DVS Gesture.}
\vspace{-0.15cm}
\begin{tabular}{ | m{0.3cm} | m{0.25cm} | m{0.25cm} | m{0.25cm} |m{0.27cm} |m{0.27cm} |m{0.27cm} |m{0.27cm} |m{0.25cm} |m{0.25cm} |m{0.25cm} |m{0.25cm} |m{0.32cm} |} 
  \hline
   \rotatebox{90}{\parbox{1.30cm}{\centering\textbf{Attack}}} & \rotatebox{90}{\parbox{1.30cm}{\centering\textbf{Hand Clap}}} & \rotatebox{90}{\parbox{1.30cm}{\centering\textbf{RH Wave}}} & \rotatebox{90}{\parbox{1.30cm}{\centering\textbf{LH Wave}}} & \rotatebox{90}{\parbox{1.30cm}{\centering\textbf{RH Clockwise}}}& \rotatebox{90}{\parbox{1.30cm}{\centering\textbf{RH \hspace{-0.05cm}Counter Clockwise}}}& \rotatebox{90}{\parbox{1.30cm}{\centering\textbf{LH Clockwise}}}& \rotatebox{90}{\parbox{1.30cm}{\centering\textbf{LH \hspace{-0.05cm}Counter Clockwise}}} & \rotatebox{90}{\parbox{1.30cm}{\centering\textbf{Arm Roll}}} & \rotatebox{90}{\parbox{1.30cm}{\centering\textbf{Air Drum}}}& \rotatebox{90}{\parbox{1.30cm}{\centering\textbf{Air Guitar}}} & \rotatebox{90}{\parbox{1.30cm}{\centering\textbf{Other}}}& \rotatebox{90}{\parbox{1.30cm}{\centering\textbf{Average}}}\\
  \hline
  \cite{spikeFool} & $\hspace{-0.1cm}90.3$ & $\hspace{-0.1cm}99$ & $\hspace{-0.1cm}89.8$& $\hspace{-0.1cm}87.3$& $\hspace{-0.1cm}79.7$& $\hspace{-0.1cm}49.7$& $\hspace{-0.1cm}51.5$& $\hspace{-0.1cm}63.6$& $\hspace{-0.1cm}79.1$& $\hspace{-0.1cm}92.3$& $\hspace{-0.1cm}64.7$ & $\textbf{77}$\\
  \hline
  This work & $\hspace{-0.1cm}100$ & $\hspace{-0.1cm}100$ & $\hspace{-0.1cm}100$& $\hspace{-0.1cm}100$& $\hspace{-0.1cm}75.8$& $\hspace{-0.1cm}86.2$& $\hspace{-0.1cm}89.7$& $\hspace{-0.1cm}100$& $\hspace{-0.1cm}66.9$& $\hspace{-0.1cm}70.5$& $\hspace{-0.1cm}75.2$ & $\textbf{87.6}$\\ 
  \hline
\end{tabular}
\label{tab:uap_comparison}
\vspace{-0.2cm}
\end{table}

Only \cite{spikeFool} proposes a solution providing results for the IBM Gesture IBM. The comparison is given in Table \ref{tab:uap_comparison} which shows the ASR per class, as well as the average ASR across all classes in the last column. As it can be seen, our attack outperforms \cite{spikeFool} in $8$ out of $11$ classes (while achieving $100\%$ ASR on $5$ classes) and on average across all classes with 87.6\% as opposed to 77\%. The perturbation size is not reported in \cite{spikeFool}, while we report it for our attack in Table \ref{tab:uap_results}. Qualitatively, our attack is more stealthy as the spike pattern of the $UAP$ is sparse and is distributed both spatially and temporally, while in \cite{spikeFool} the patch is limited to the area where the actual gesture is performed making it noticeable.
\section{Conclusion} \label{sec:conclusion}

We introduced two innovative adversarial attack methods for SNNs leveraging spatiotemporal gradients in the spiking domain: an input-specific attack and a universal adversarial attack. The input-specific attack generates adversarial examples from any dataset sample while introducing minimal, imperceptible perturbations. The universal adversarial attack crafts a single reusable perturbation that significantly degrades accuracy across the vast majority of inputs, making it highly efficient for real-time deployment. Results on SNNs trained for the NMNIST and IBM DVS Gesture vision datasets, show that the proposed attacks outperform across all evaluated metrics all prior state-of-the-art attacks. Furthermore, for the first time we demonstrate adversarial attacks on the SHD neuromorphic auditory dataset. Overall, this work provides new insights into adversarial robustness of SNNs and highlights critical security implications for neuromorphic computing, emphasizing the need for dedicated defense mechanisms tailored to SNNs. Existing defenses, such as adversarial training \cite{BaSiRa18, spikeFool} and input filtering \cite{marchisio2021dvsattacksadversarialattacksdynamic}, offer potential countermeasures but often introduce trade-offs like increased computational overhead or reduced accuracy on clean inputs. As future work, we aim to develop SNN-specific defense mechanisms that effectively balance robustness, efficiency, and real-world applicability.

\bibliographystyle{IEEE}

\end{document}